\documentstyle[12pt]{article}

\newcommand{\be}{\begin{equation}}
\newcommand{\ee}{\end{equation}}
\newcommand{\ba}{\begin{eqnarray}}
\newcommand{\ea}{\end{eqnarray}}
\newcommand{\spz}{\hspace{0.7cm}}
\newcommand{\virg}{\spz,\spz}
\newcommand{\pvirg}{\spz;\spz}

\newcommand{\la}{\lambda}
\newcommand{\dx}{\partial_x}
\newcommand{\dt}{\partial_t}

\newcommand{\tp}{\otimes}

\newcommand{\lt}{\left(}
\newcommand{\rt}{\right)}

\newcommand{\bC}{\mbox{\bf C}}

\newcommand{\buno}{\mbox{\bf 1}}

\newcommand{\cL}{{\cal L}}

\newcommand{\cN}{{\cal N}}

\newcommand{\cP}{{\cal P}}

\newcommand{\cV}{{\cal V}}
\newcommand{\cW}{{\cal W}}

\newcommand{\NP}[1]{Nucl.\ Phys.\ {\bf #1}}
\newcommand{\PL}[1]{Phys.\ Lett.\ {\bf #1}}

\newcommand{\CMP}[1]{Comm.\ Math.\ Phys.\ {\bf #1}}
\newcommand{\CPAM}[1]{Comm.\ Pure\ Appl.\ Math.\ {\bf #1}}

\newcommand{\PRL}[1]{Phys.\ Rev.\ Lett.\ {\bf #1}}

\begin{document}
\sloppy
\renewcommand{\thefootnote}{\fnsymbol{footnote}}

\newpage
\setcounter{page}{1}

\vspace{0.7cm}
\begin{flushright}
DTP-98-17\\
PM/98-09\\
February 1998
\end{flushright}
\vspace*{1cm}
\begin{center}
{\bf On the Null-Vectors in the Spectra of the 2D Integrable Hierarchies}
        \footnote{Work supported  by European Union under contract FMRX-CT96-0012} \\
\vspace{1.8cm}
{\large D.\ Fioravanti $^a$ and M.\ Stanishkov $^b$ \footnote{On leave from I.N.R.N.E. - Sofia, Bulgaria\\ 
E-mail: Davide.Fioravanti@durham.ac.uk, marian@lpm.univ-montp2.fr}}\\
\vspace{.5cm}
$^a${\em Department of Mathematical Sciences - Univ. of Durham\\
     South Road, DH1 3LE DURHAM, England} \\
$^b${\em Laboratoire de Physique Mathematique - Univ. de Montpellier II\\
     Pl. E. Bataillon, 34095 Montpellier, France} \\
\end{center}
\vspace{1cm}

\renewcommand{\thefootnote}{\arabic{footnote}}
\setcounter{footnote}{0}

\begin{abstract}
{\noindent We propose} an alternative description of the spectrum 
of local fields in the classical limit of the integrable quantum field 
theories. It is close to similar constructions used in the geometrical treatment of
${\cW}$-gravities \cite{SS}. Our approach provides a systematic way of deriving the 
null-vectors that appear in this construction. We present explicit results for the case of 
the $A_1^{(1)}$-(m)KdV and the $A_2^{(2)}$-(m)KdV hierarchies, different classical limits
of 2D CFT's. In the former case our results coincide with the classical limit 
of the construction \cite{BBS}. Some hints about quantization and off-critical 
treatment are also given.
\end{abstract}
\newpage

\section{Introduction}
One of the most important open problems in the 
two dimensional integrable 
quantum field theories ({\bf 2D-IQFT's}) is the 
construction of the spectrum of the local fields and consequently the 
computation of their correlation functions. These problems are
somewhat 
connected to one another within the simplest integrable theories --  {\bf 2D-CFT's} \cite{BPZ} -- which have 
a covariance under the infinite dimensional Virasoro 
symmetry as their common crucial property . In fact the Verma modules of this algebra classify all the local 
fields in 2D-CFT's and turn out to be reducible owing to the occurrence of 
vectors of null hermitian product with all other vectors, the so called 
null-vectors. Factorization by the modules generated over the 
null-vectors leads to a number of very interesting algebraic-geometrical 
properties such as  fusion algebras, differential equations for 
correlation function, etc. .

Unfortunately this beautiful picture collapses when one pushes the system away
from criticality by perturbing the original CFT with some relevant local field
 $\Phi$:
\be
S=S_{CFT}+\mu\int d^2z\, \Phi(z,\bar{z}).
\label{pert}
\ee     
Of the infinite dimensional Virasoro symmetry only the Poincar\'e subalgebra
survives the perturbation. Actually, the CFT possesses a bigger ${\cW}$-like 
symmetry and in particular it is invariant under an infinite dimensional 
abelian subalgebra of the latter \cite{SY}. With suitable deformations, this 
abelian subalgebra survives the perturbation (\ref{pert}), resulting in the 
so-called  local integrals of motion ({\bf LIM}). Being 
abelian, this symmetry does not carry sufficient information, and in particular
one cannot build the spectrum of a theory of type (\ref{pert}) by means of 
{\bf LIM} alone.

It was conjectured in \cite{BBS} that one could add to {\bf LIM} $I_{2m+1}$ 
additional  charges $J_{2m}$ in such a way that the 
resulting algebra would be sufficient to create all the states of the 
perturbed theory (\ref{pert}). Therein it was also discovered that a sort of 
null-vector condition appears in the above procedure leading, as in CFT's, 
to certain equations for the form factors. However, what remains unclear in
 \cite{BBS} is the {\it general procedure} of finding the null-vectors, and in 
particular the {\it symmetry structures} lying behind them. Besides, heavy use is
made of the very specific form of the form factors of the RSG model, and it is 
not clear how to extend this procedure to other integrable theories .

Another possible treatment of the spectrum of CFT's is
connected to the so called spinon fields, essentially the chiral vertex 
operators. It was shown in the context of Wess-Zumino-Witten models that this 
approach gives an appropriate basis for the representations of the 
corresponding Yangian symmetry. It is very likely that all this will prove 
useful in other CFT's as well and will possibly be extendible outside the 
critical point,
thus providing a connection with quantum groups and Yangian symmetries of 
integrable theories of type (\ref{pert}). Indeed, as we shall see below, the
classical limit of the elementary vertex operators appears in the regular expansion of our basic object, 
the transfer matrix $T(x;\la)$.  

In this article we propose a systematic treatement of the above problems which
 is essentially a variation of the  Inverse Scattering Method ({\bf ISM}) 
\cite{Fa84}. The central idea is to base the whole construction on the
so-called 
Dressing Symmetry Transformations \cite{Sem} connecting these to the
 usual
 way of finding integrable systems \cite{DS}. In fact, our basic objects 
will be the  transfer matrix $T(x;\la)$, which generates the dressing, 
and
the  resolvents $Z^X(x;\la)$, the dressed generators of the underlying 
algebra. In this article we are mainly concerned with the semiclassical limit of CFT's 
\cite{Ge,BLZ1,FRS}, namely the $A_1^{(1)}$-KdV and the $A_2^{(2)}$-KdV 
systems \cite{DS}. In compliance with the basic work \cite{BLZ1}, one  quantizes 
the corresponding mKdV system by replacing the Kac-Moody algebra with the corresponding 
deformed algebra $U_q(A_1^{(1)})$ or  $U_q(A_2^{(2)})$ and the mKdV field $\phi$ with the Feigin--Fuchs 
free field \cite{FF}. As explained in \cite{FRS}, the importance of 
considering also the $A_2^{(2)}$-mKdV hierarchies is due to the fact that the 
quantization of this second semiclassical system exhausts the integrability 
directions
of theories of type  (\ref{pert}), starting from minimal models.

Besides, we assume  the monodromy matrix for the perturbed theory (\ref{pert})
, according to \cite{BLZ3}, to be 
\be
{\bf T}(x,\bar{x};\la;\mu) =\bar{T}(\bar{x};\mu/\la) T(x;\la),
\ee 
where the precise definition of the anti-chiral monodromy matrix $\bar{T}(\bar{x};\la)$ will be given by formula (\ref{Tba}).
In the sequel we shall only mention the off-criticality construction briefly,
setting the topic aside for future publications \cite{35,36}

\section{The $A_1^{(1)}$--mKdV: the regular expansion of the transfer matrix}
As already observed in  \cite{Ge}, the classical limit ($c\to
-\infty$) of CFT's is described by the second Hamiltonian structure of the
(usual) KdV which corresponds to $A_1^{(1)}$ in the Drinfeld-Sokolov
scheme \cite{DS}.The KdV variable  $u(x,t)$ is related to the mKdV variable
$v(x,t)$  by the Miura transformation \cite{Miu} 
$u=-v^2 +v'$, which is the classical
counterpart of the Feigin-Fuchs transformation \cite{FF}.
In fact the mKdV equation is
\be
\dt v=-\frac{3}{2}v^2v'-\frac{1}{4}v'''.
\label{mkdv}
\ee
and the mKdV field $v=-\phi'$ is the derivative of the Darboux field $\phi$. 
The equation (\ref{mkdv}) can be re-written \cite{Lax} as a null curvature
condition $[ \dt - A_t , \dx - A_x ] = 0 $ for connections
belonging to the $A_1^{(1)}$  loop  algebra 
\ba
A_x &=& - v h + (e_0 + e_1),  \nonumber \\
A_t &=& \la^2(e_0 + e_1 - vh) -
\frac{1}{2}[(v^2-v')e_0 + (v^2+v')e_1] - \frac{1}{2}(\frac{v''}{2}-v^3)h
\label{lax}
\ea
where the generators $e_0,e_1,h$  are chosen in the
fundamental representation and canonical gradation 
of the $A_1^{(1)}$ loop algebra
\be
e_0=\la E=\left(\begin{array}{cc} 0 & \la \\
                                  0 & 0 \end{array}\right)\virg
e_1=\la F=\left(\begin{array}{cc} 0 & 0 \\
                                  \la & 0 \end{array}\right)\virg
h  = H=\left(\begin{array}{cc} 1 & 0 \\
                               0 & -1 \end{array}\right) .
\label{gen}
\ee
A remarkable geometrical interest is obviously attached to 
the transfer matrix which performs the parallel transport 
along the $x$-axis, and is thus the solution of the associated linear problem
\be
\dx T(x;\la) = A_x(x;\la)T(x;\la)  . 
\label{T1}
\ee
The formal solution of the previous equation is given by
\be
T(x,\la) =e^{H\phi(x)}{\cal P}
\exp\lt\la \int_0^xdy
(e^{-2\phi(y)} E+ e^{2\phi(y)} F ) \rt =\left(\begin{array}{cc} A & B \\
                                                                C & D \end{array}\right) ,
\label{forsol}
\ee
where the expansions of the entries are
\ba
A(x;\la) &=&e^{\phi(x)}+O(\la^2) \virg  B(x;\la)=\la e^{\phi(x)} \int_0^x dy e^{-2\phi(y)}+O(\la^3),
\nonumber \\
C(x;\la) &=& A(-\phi(x)) \virg D(x;\la)=B(-\phi(x)) .
\label{ABCD}
\ea
Note that the first terms of the expansion (\ref{ABCD}) are exactly the
classical limits of the two {\it elementary} vertex operators \cite{MS}. 
Besides, the expression (\ref{forsol}) defines $T(x,\la)$ as an entire 
function of $\la$ with an essential singularity at $\la=\infty$ where it is governed 
by the corresponding asymptotic expansion. The two expansions give rise to different algebraic 
and geometric structures, as we shall see below. The regular expansion 
is typically employed  in the derivation of Poisson-Lie 
structures for Dressing Symmetries \cite{Sem}. 
The second type of expansion plays instead a  crucial role 
in obtaining the local integrals of motion \cite{Fa84,DS}. 
In this article we shall only give an overview of how the aforementioned
approaches merge actually into a single procedure which, however,
produces two different kinds of symmetries \cite{35}.

In our case the formal solution (\ref{forsol}) can be expressed  as
an expansion in positive powers of $\la$ with an infinite  
radius of convergence and non-local coefficients:
\be
T(x;\la)=e^{H\phi(x)} \sum_{k=0}^{\infty}
\la^k \int_{x\geq x_1 \geq x_2 \geq ... \geq
x_k\geq 0}K(x_1)K(x_2)...K(x_k) dx_1 dx_2 ... dx_k
\label{regexp}
\ee
where $K(x)= e^{-2\phi(x)}E+e^{2\phi(x)}F$. 
After calculating the expression (\ref{regexp}) for $x=L$  and taking
the 
trace, we obtain the regular expansion for $\tau(\la)=tr{T(L;\la)}$ in terms of the non-local 
conserved charges in involution. However, one may obtain a larger number 
of non-local conserved charges not in 
involution, i.e. so that the charges commute with local
hamiltonian of the mKdV (\ref{mkdv}) but not between themselves. This can 
be carried out by means of the dressing techniques in the following way.
Let $X=H,E,F$ be one of the generators (\ref{gen}). 
By assuming the regular expansion (\ref{regexp}), let us construct the
generic  resolvent by dressing
\be 
Z^X(x,\la)=(TXT^{-1})(x,\la)=\sum_{k=0}^\infty \la^k Z^X_k  . 
\label{dres}
\ee
$Z^X(x,\la)$ is clearly a  resolvent for the operator ${\cL}=\dx -
A_x$   (\ref{lax}) since by construction it satisfies 
\be
[{\cL},Z^X(x;\la)]=0  .
\label{resdef}
\ee
The foregoing property of the  resolvent justifies the conclusion that, 
once the quantity is determined
\be
\Theta^X_n(x;\la)=(\la^{-n}Z^X(x;\la))_-=\sum_{k=0}^{n-1} \la^{k-n} Z^X_k  ,
\label{theta}
\ee 
the commutator $[{\cL},\Theta^X_n(x;\la)]$ is of degree zero in $\la$.
Therefore it is possible to construct the gauge transformation  
\ba
\delta^X_n A_x &=& -\delta^X_n{\cL} = -[\Theta^X_n(x;\la),{\cL}] , \nonumber \\
\delta^X_n A_t &=& -[\Theta^X_n(x;\la), \dt - A_t],
\label{gautra}
\ea
which preserves the zero curvature condition by construction. It will also be 
a true symmetry of the model in case the last term in (\ref{gautra}) is
proportional to $H$:
\be
\delta^X_n A_x=H \delta^X_n  \phi' .
\label{var}
\ee
This depends, for X fixed, on whether n is even or
odd. Indeed, by directly substituting the regular
expansion (\ref{regexp}) in (\ref{dres}), one can obtain :  
\ba
Z^H_{2m}(x)=a^H_{2m}(x)H  \virg   Z^H_{2m+1}(x)=b^H_{2m+1}(x)E+c^H_{2m+1}(x)F \nonumber\\
Z^E_{2n}(x)=b^E_{2n}(x)E+c^E_{2n}(x)F  \virg  Z^E_{2n+1}(x)=a^E_{2n+1}(x)H  \nonumber \\
Z^F_{2p}(x)=b^F_{2p}(x)E+c^F_{2p}(x)F \virg  Z^F_{2p+1}(x)=a^F_{2p+1}(x)H  .
\label{Z's}
\ea
The variation (\ref{gautra}) can be explicitly calculated as 
\be
\delta^X_n A_x = [Z^X_{n-1},E+F] 
\label{expvar}
\ee
and hence it is clear that $Z^X_{n-1}$ cannot
contain any term proportional to $H$. 
The conclusions are that
\begin{itemize}
\item in the $\Theta^H_n$ case $n$, in
(\ref{gautra}), must be even,
\item in the  $\Theta^E_n$ and  $\Theta^F_n$ case $n$ must conversely be odd. 
\end{itemize}
Besides, it is possible to show  by direct calculation  that these
infinitesimal transformation generators form a representation of  a (twisted) Borel subalgebra $A_1\tp \bC$, (of the loop algebra $A_1^{(1)}$)
\be
[\delta^X_m , \delta^Y_n] = \delta^{[X,Y]}_{m+n} \pvirg X,Y = H,E,F  .
\label{delalg}
\ee
The first generators of this algebra are explicitly given by: 
\ba
\delta^E_1 \phi'(x) &=&  e^{2\phi(x)}          \nonumber  \\
\delta^F_1 \phi'(x) &=&  - e^{-2\phi(x)}    \nonumber  \\
\delta^H_2 \phi'(x) &=& e^{2\phi(x)} \int_0^x dy e^{-2\phi(y)} +  e^{-2\phi(x)} \int_0^x dy 
e^{2\phi(y)}
\label{che}
\ea
and the rest are derived from these by commutation.
The N\"other currents $J^X_{t,n}$, originating from this symmetry algebra, 
can easily be found by applying the transformations (\ref{expvar}) 
to both members of the continuity equation (\ref{mkdv}). 
To the $J^X_{t,n}$ correspond the non-local charges 
\ba
Q^H_{2m+2}&=&\int_0^L J^H_{t,2m+2}=a^H_{2m+2}(L)   \nonumber \\
Q^E_{2n+1}&=&\int_0^L J^E_{t,2n+1}=a^E_{2n+1}(L)  \nonumber \\
Q^F_{2p+1}&=&\int_0^L J^F_{t,2p+1}=a^F_{2p+1}(L)
\label{char}
\ea
which are not necessarily conserved (depending on the boundary conditions), 
due to non-locality. Note that the action by which these
charges generate the transformations (\ref{expvar}),(\ref{che}) is not always symplectic.
Indeed, the following Poisson brackets   
\ba
\delta^E_1v=\{Q^E_1,v\} \virg  \delta^F_1v=\{Q^F_1,v\}  \nonumber \\
\delta^H_2v=\{Q^H_2,v\} + Q^E_1\{Q^F_1,v\} - Q^F_1\{Q^E_1,v\} 
\label{poilie}
\ea
denote how in the first case the action is symplectic, while in the second it
is of Poisson-Lie type \cite{Sem}.
It is possible to verify by explicit calculation  that the charges 
themselves close a (twisted) Borel subalgebra $A_1\tp \bC$, 
(of the loop algebra $A_1^{(1)}$).

\section{Asympotic expansion of the transfer matrix}
Around the point $\la=\infty$ the system is governed by the asymptotic 
expansion. It can be obtained through the procedure described in  \cite{Fa84,DS}. The main
idea is to apply a gauge transformation $S(x)$ on the Lax operator ${\cL}$
in such a way that its connection lie in the corresponding Cartan subalgebra:
\be
(\dx - A_x(x))S(x)=S(x)(\dx+D(x))
\label{diag}
\ee
where $D(x;\la)=d(x;\la)H$. Because of the previous equation
 $T(x;\la)$ has the form 
\be 
T(x;\la)=KG(x;\la)e^{-\int_0^x dy D(y)}
\label{asyexp1}
\ee
where we put $S=KG$ with 
\be
K =\frac{\sqrt{2}}{2}\left(\begin{array}{cc}  1 & 1 \\
                                              1 & -1 \end{array}\right),
\ee
while $G$ verifies the following equation (of 
Riccati type for the off-diagonal part of $G$)
\be
\dx G + \tilde{A}_x G=GD  \virg    \tilde{A_x}=K^{-1}A_xK .
\label{ricc}
\ee
It is clear now that the previous equation can be solved by finding the 
asymptotic expansion for $d(x;\la)$ and expressing the asymptotic expansion of
 $G(x;\la)$ in terms of off-diagonal matrices 
\be
d(x;\la)=\sum_{k=-1}^{\infty}\la^{-k}d_k(x) \virg G(x;\la)=\buno +
\sum_{j=1}^{\infty}\la^{-j}G_j(x)  
\label{asyexp2} 
\ee 
where the $G_j(x)$ matrices are off-diagonal with entries $(G_j(x))_{12}=g_j(x)$ and
$(G_j(x))_{21}=(-1)^{j+1}g_j(x)$.  The explicit recurrence formulas for the
$g_j(x)$ and the $d_k(x)$ are:  \ba g_{j+1}=\frac{1}{2}(g'_j + v
\sum_{k=1}^{j-1}g_{i-j}g_j) \nonumber \\ d_{-1}=-1,d_0=0; \spz j>0,
d_j=(-1)^{j+1} vg_j \label{asyexp3} \ea Note that the $d_{2n}(x)$ are
exactly the charge densities (of the mKdV equation) resulting from the
asymptotic expansion of $\tau(\la)=trT(\la)$. 

It is likewise known \cite{DS} that the construction of the mKdV flows goes
through the definition of a  resolvent $Z(x;\la)$ defined through the
following property of its asymptotic expansion 
\be 
[{\cL},Z(x;\la)]=0,\spz
Z(x,\la)=\sum_{k=0}^\infty \la^{-k} Z_k, \spz Z_0=E+F .  
\label{asyres} 
\ee
From the definition (\ref{asyres}) it is
possible to derive how the  resolvent $Z$ is obtained by {\it
dressing} from the asymptotic expansion of $T$ (\ref{asyexp1}),(\ref{asyexp2}),(\ref{asyexp3}).  
\be
Z(x,\la)=(THT^{-1})(x,\la);  
\label{asydre} 
\ee
or for the
modes of the $\la$-expansion \be Z_{2k}(x)=b_{2k}(x)E+c_{2k}(x)F \virg
Z_{2k+1}(x)=a_{2k+1}(x)H, 
\label{solasy} 
\ee 
where for example 
\ba
a_1 &=& -v \virg a_3=\frac{1}{4}v^3 - \frac{1}{8} v'' \nonumber \\
b_2 &=& \frac{1}{4}v^2 + \frac{1}{4} v' \virg  b_4=-\frac{3}{16}v^4 + \frac{1}{8} v''v- \frac{1}{16} v'^2 - \frac{3}{8} v'v^2 + \frac{1}{16} v''', etc.     . 
\label{ab} 
\ea 
Note that the $b_{2k}$ are exactly the charge densities of the KdV
equation (and are thus expressible in terms of the $u(x)$ field only). They
differ by a total derivative from the $d_{2k}(x)$. The $a_{2k+1}$ instead
are not expressible in terms of $u(x)$ alone, and have to be obtained from
$a_1=-v$. As in the regular case, the system enjoys a gauge symmetry 
of the form (\ref{gautra}) with the constraint (\ref{var}). It turns 
out that this gauge transformation coincides 
with the commuting mKdV flows:  
\be \delta_{2k+1}A_x =-[\theta_{2k+1}(x;\la),{\cL}], 
\ee 
where the $\theta_{2k+1}$ are the Lax
connections associated to $A_x$ 
\be
\theta_{2k+1}(x;\la)=(\la^{2k+1}Z(x;\la))_+=\sum_{j=0}^{2k+1}
\la^{2k+1-j} Z_j(x) .  
\ee 
Let us set aside for
\cite{35} the construction of the flows deriving from $Z^E=TET^{-1}$ and $Z^F=TFT^{-1}$ and
no more commuting with the $\delta_{2k+1}$ of the hierarchy, but rather
closing with them a spectrum generating  algebra, and let us
concentrate our attention on the phase spaces of mKdV and KdV systems, i.e. 
those objects which at the quantum level constitute the spectrum of local fields.

Our approach to the spectrum of local fields is close to similar constructions used 
in the geometrical treatment of ${\cW}$-gravities \cite{SS}. It was proposed there 
to start from the corresponding Frenet equations and to treat the entries 
of the resulting gauge connection as independent variables. Similarly, we propose 
here to treat the components of $Z$ as  fundamental fields. Unlike 
the case \cite{SS}, infinitely many of them appear here because in 
this case the Lax connections belong to the $A_1^{(1)}$ loop algebra.

Let us start by considering the $a_{2n+1}$, $b_{2n}$, and $c_{2n}$ fields of (\ref{solasy}).
The differential equation (\ref{asyres}) tells us that not all of
them are independent. We may for example express the $c_{2n}$ in terms of the
{\it basic} fields $b_{2n}$ and $a_{2n+1}$.
Thus, for the mKdV hierarchy the Verma module ${\cV}^{mKdV}_{{\bf 0}}$ of the identity 
will  be freely generated by the repeated action of the $\delta_{2k+1}$
on the polynomials in the $b_{2n}$ and the $a_{2k+1}$ 
\be
{\cV}^{mKdV}_{{\bf 0}}=\{\delta_{2k_1+1}\delta_{2k_2+1}\dots \delta_{2k_M+1}{\cP}(b_2,b_4,
\dots ,b_{2N},a_1,a_3,\dots ,a_{2P+1})\} .  
\label{veridm}
\ee
But not all of these fields are independent yet. Instead, it happens 
that they obey a number of additional constraints. As a consequence 
some of their linear combinations are zero, giving a sort of ''classical null-vectors ''.  
The first constraint 
\be
Z^2=\buno \spz i.e. \spz \sum_{i=0}^nb_{2n-2i}(b_{2i}+a'_{2i-1})+\sum_{i=0}^{n-1}a_{2n-2i-1}a_{2i+1} = 0
\label{con}
\ee
originates from the dressing relation with the transfer 
matrix $T$ (\ref{asydre}).The second one is just {\it the equation of motion} of 
$Z$ \cite{DS}
\be
\delta_{2k+1}Z=[\theta_{2k+1},Z]
\label{eqsmot}
\ee
(which may easily be deduced from (\ref{asyres})) wich  allows us to establish the
action of the $\delta_{2k+1}$ on these fields
\ba
\delta_{2k+1}a_{2n+1}=\sum_{i=0}^{n}(a'_{2n+2k-2i+1}b_{2i}-a'_{2i-1}b_{2m+2k-2i+2})        , \nonumber \\  
\delta_{2k+1}b_{2n}=2\sum_{i=0}^{n-1}(a_{2n+2k+1}b_{2n+2k-2i}-a_{2n+2k-2i+1}b_{2i})   
\label{deltaab}
\ea
Actually, the whole set of all the null vectors ${\cN}^{mKdV}_{{\bf 0}}$ is 
obtained by the successive application of (\ref{con}),(\ref{eqsmot}) and their 
linear combinations.
In conclusion, the family of 
the identity $[{\bf 0}]^{mKdV}$ of the mKdV hierarchy is 
obtained as a factor space:
\be 
[{\bf 0}]^{mKdV}={\cV}^{mKdV}_{{\bf 0}} / {\cN}^{mKdV}_{{\bf 0}}.
\label{conidm}
\ee
On the other hand we can make two observations:
\begin{enumerate}
\item the relation (\ref{con}) can be re-written in terms of the $b_{2k}$
alone
\be
b_{2n}+\sum_{i=1}^n[b_{2n-2i}b_{2i}-2b_2b_{2n-2i}b_{2i-2}-
\frac{1}{2}b_{2n-2i}b'_{2j-2}+\frac{1}{4}b'_{2n-2j}b'_{2j-2}] = 0 ;
\label{constb}
\ee  
\item also the variation of $b_{2n}$ in (\ref{deltaab}) can be written  in
terms of the $b_{2k}$ alone, using the relationships (\ref{asyres}) between 
$a_{2k+1}$ and $b_{2k}$
\be
\delta_{2k+1}b_{2n}=\sum_{i=0}^{n-1}(b'_{2n+2k-2i}b_{2i}-b'_{2i}b_{2m+2k-2i}).
\label{deltab}
\ee
\end{enumerate}
Therefore for the KdV equation the Verma module  ${\cV}^{KdV}_{{\bf 0}}$ of the 
identity will be freely generated by the repeated action of the
$\delta^H_{2k+1}$ on the polynomials only in the $b_{2n}$: 
\be
{\cV}^{KdV}_{{\bf 0}}=\{\delta^H_{2k_1+1}\delta^H_{2k_2+1} \dots \delta^H_{2k_M+1}{\cP}(b_2b_4\dots b_{2N})\} .
\label{veridk}
\ee
It turns out to be a sort of {\it reduction} of the 
Verma module ${\cV}^{mKdV}_{{\bf 0}}$ (\ref{veridm}) of the mKdV hierarchy.
The whole set of null vectors ${\cN}^{KdV}_{{\bf 0}}$ is given by successive 
applications of (\ref{deltab}) and  (\ref{constb}). 
As for the mKdV case the ({\it conformal}) family of the identity 
$[{\bf 0}]^{KdV}$ of the KdV hierarchy is obtained as a factor space 
of ${\cV}^{KdV}_{{\bf 0}}$  :
\be 
[{\bf 0}]^{KdV}={\cV}^{KdV}_{{\bf 0}} / {\cN}^{KdV}_{{\bf 0}}.
\label{conidk}
\ee
Therefore we are led to the same scenario which arise
also in the classical limit of the construction \cite{BBS}. Nevertheless, in our 
approach the generation of null-vectors is automatic (see equations  (\ref{deltab}) 
and  (\ref{constb})). In addition , our approach is applicable to any other 
integrable system, based on a Lax pair formulation. We shall demonstrate this below
by using the example of the  $A_2^{(2)}$-mKdV system.

Other local fields of the mKdV system are the primary fields, i.e. 
the exponentials $e^{m\phi}, m=0,1,2,3,\dots$ of the bosonic field. 
Indeed, for $m=0$ we obtain just the identity $\buno$, the 
elementary primary field $e^{\phi}$ (m=1) appears in the 
regular expansion (\ref{ABCD}) of the transfer matrix $T(x;\la)$ and the 
other primary field $e^{m\phi}, m>1$ are the ingredients of regular expansion 
of the power $T^m(x;\la)$. The previous construction of the identity operator 
family suggests the following form for the Verma module  ${\cV}^{KdV}_m$ of 
the primary $e^{m\phi}, m=0,1,2,3,\dots$:  
\be
{\cV}^{KdV}_m=\{\delta_{2k_1+1}\delta_{2k_2+1}\dots \delta_{2k_M+1}[{\cP}(b_2,b_4,\dots ,b_{2N})e^{m\phi}]\} .  
\label{vermpm}
\ee
As for the identity family, we have to subtract all the null-vectors  (\ref{deltab}) 
and  (\ref{constb}). Besides, in this case, we have to take into account
the null-vectors coming from the equations of motion of the power $T^m(x;\la)$ 
\be
\delta_{2k+1}T^m=\sum_{j=1}^mT^j\theta_{2k+1}T^{m-j}.
\label{eqmT^m}
\ee
By successive applications of   (\ref{deltab}), (\ref{constb}) and (\ref{eqmT^m}) 
we obtain the whole
null-vector set ${\cN}^{KdV}_{{\bf m}}$. In conclusion the spectrum is again the 
factor space 
\be
[{\bf m}]={\cV}^{KdV}_m / {\cN}^{KdV}_{{\bf m}} .
\label{conmk}
\ee

\section{The $A_2^{(2)}$-KdV.}
Let us show that our approach is easily applicable to other integrable systems. Here we 
consider the case of the $A_2^{(2)}$-KdV equation. The reason is that it can be considered
as a different classical limit of the {\bf CFT's} \cite{FRS}. Consider the matrix 
representation of the $A_2^{(2)}$-KdV equation:
\be
\dt{\cL}=[{\cL}, A_t]
\label{mKdV2}
\ee
where
\be
{\cL}=\dx-A_x \virg A_x= \phi' h + (e_0 + e_1), \spz u(x)=-\phi'(x)^2 -\phi''(x)
\ee
and 
\be
e_0= \left(\begin{array}{ccc} 0 & 0 & \la \\
                                  0 & 0 &  0  \\                                      
                                  0 & 0 &  0   \end{array}\right)\virg
e_1= \left(\begin{array}{ccc} 0 & 0 &  0 \\
                                  \la & 0 &  0  \\
                                  0 & \la & 0 \end{array}\right)\virg
h  = \left(\begin{array}{ccc} 1 & 0 & 0 \\
                               0 & 0 & 0 \\
                               0 & 0 & -1 \end{array}\right) 
\label{gen2}
\ee
are the generators of the Borel subalgebra of $A_2^{(2)}$ and $A_t$ is a 
certain connection that can be found for exemple in \cite{DS}. As shown before,
 central role is played by the transfer matrix, wich is a solution of the associated linear 
problem $(\dx  - A_x(x;\la))T(x;\la)=0$. The formal solution is in this case 
\be
T(x,\la) =e^{h\phi(x)}{\cal P}
\exp\lt \int_0^xdy
(e^{-2\phi(y)} e_0+ e^{\phi(y)} e_1 ) \rt  .
\label{trama2}
\ee
The equation (\ref{trama2}) defines $T$ as an entire function of $\la$ with an essential
singularity at $\la=\infty$. In this article we are concerned only with the corresponding 
asymptotic expansion leaving the general treatment of the problem to a further 
publication \cite{36}. As for the $A_1^{(1)}$ case, this asymptotic expansion is easily 
written by following the general procedure \cite{DS}. The result is:
\be 
T(x;\la)  = \left(\begin{array}{ccc} 1 & h^+_1 &  h^+_2 \\
                               0 &  1+h^0_3 &  h^0_1 \\
                               0 &  h^-_2 &  1+h^-_3 \end{array}\right)
\exp \lt -\int_0^x\sum_{i=0}^\infty f_i \Lambda^{-i} \rt  .
\ee
where $h^{\pm}_i,h^0_i$ are certain polinomials in $\phi'$, $f_{6k},f_{6k+2}$ are the 
densities of the local conserved charges of the $A_2^{(2)}$-KdV and $\Lambda=e_0+e_1$. Complying 
with our approach let us introduce the resolvents 
\ba
Z_1=T\Lambda T^{-1} ; \nonumber \\
Z_2=T\Lambda^2 T^{-1}
\label{res2}
\ea
satisfying as before the equations  $[\dx - A_x ,Z_i(x;\la)]=0, \spz i=1,2$ :
\ba
&Z_1& = \left(\begin{array}{ccc}     \la^{-1}a_1^{(1)}+\la^{-4}a_4^{(1)}+ \dots & \la^{-2}b_2^{(1)}+\la^{-5}b_5^{(1)}+\dots &  
1+\la^{-6}b_6^{(1)}+\dots \\
                                 1+\la^{-3}c_3^{(1)}+ \dots  &  -2\la^{-4}a_4^{(1)}+ \dots & \la^{-2}b_2^{(1)}-\la^{-5}b_5^{(1)}
+\dots  \\                                      
                                 \la^{-2}c_2^{(1)}+\dots    &  1-\la^{-3}c_3^{(1)}+\la^{-6}c_6^{(1)}+\dots  & -\la^{-1}a_1^{(1)}+\la^{-4}a_4^{(1)}+\dots  \end{array}\right) ,  \nonumber \\
&Z_2& = \left(\begin{array}{ccc}     \la^{-2}a_2^{(2)}+\la^{-5}a_5^{(2)}+\dots  & 1+\la^{-3}b_3^{(2)}+\la^{-6}b_6^{(2)}+\dots &  
\la^{-4}b_4^{(2)}+\dots \\
                                 \la^{-1}c_1^{(2)}+\la^{-4}c_4^{(2)} \dots  &  -2\la^{-2}a_2^{(2)}+ \dots & 1-\la^{-3}b_3^{(2)}-\la^{-6}b_6^{(2)}
+\dots  \\                                      
                                 1+\la^{-6}c_6^{(2)}+\dots    &  -\la^{-1}c_1^{(2)}+\la^{-4}c_4^{(2)}+\dots  & \la^{-2}a_2^{(2)}-\la^{-5}a_5^{(2)}+\dots  \end{array}\right) \nonumber .
\ea
For example, some expressions for the fiels in the entries of $Z_i, i=1,2$, as a function of $v$ 
and its derivative, are:
\ba
a_1^{(1)} &=& -v ; \nonumber \\
b_2^{(1)} &=& - \frac{1}{3} v^2 + \frac{1}{3} v' \virg  c_2^{(1)}=- \frac{1}{3} v^2 - \frac{2}{3} v' ; \nonumber \\
c_3^{(1)} &=& \frac{1}{3} v^3 + \frac{1}{3} vv' - \frac{1}{3} v'' \virg  b_3^{(2)}=-\frac{2}{3} vv' + \frac{1}{3} v'' ; \nonumber \\
b_4^{(2)} &=& -\frac{1}{9} v^4 + \frac{1}{9} v'^2 - \frac{2}{9} vv''+ \frac{2}{9} v'v^2 - \frac{1}{9} v''', etc.  .
\ea
The equation (\ref{mKdV2}) is invariant under a gauge transformation of the form
(\ref{gautra}) . This latter will be a true symmetry  provided the variation is proportional to
h : $\delta A_x= h \delta \phi'$. We construct the appropriate gauge parameters by
means of the resolvents (\ref{res2}) in a way similar to what we did in the $A_1^{(1)}$-KdV case:
\be
\theta_{6k+1}(x;\la)=(\la^{6k+1}Z_1(x;\la))_+  \virg   \theta_{6k-1}(x;\la)=(\la^{6k-1}Z_2(x;\la))_+
\label{theta's}
\ee
which results in the following transformations for the  $A_2^{(2)}$-mKdV field
\be
\delta_{6k+1}\phi'=\dx a_{6k+1}^{(1)} \virg   \delta_{6k-1}\phi'=\dx a_{6k-1}^{(2)}  .
\label{var2}
\ee
One can easily recognize in (\ref{var2}) the infinite tower of the commuting
$A_2^{(2)}$-mKdV flows.

Now, in accordance with our geometrical conjecture, we would like to treat the entries of
the transfer matrix $T$ and of the resolvents $Z_i, i=1,2$ as independent fields 
and to build the spectrum of the local fields of $A_2^{(2)}$-KdV by means of 
them alone. As in the $A_1^{(1)}$ case, it turns out that not all of them are 
independent. If the defining relations of the resolvents are used, it is easy to see, that 
the entries of the lower triangle of both $Z_i$ can be expressed in terms of the rest.
Therefore, taking also into account the gauge symmetry of the system, one is led 
to the following proposal about the construction of the Verma module of the identity: 
\be
{\cV}^{mKdV}_{{\bf 0}}=\{\delta_{6k_1+1}\dots \delta_{6k_M+1}\delta_{6l_1-1}\dots \delta_{6l_N-1}
{\cP}(b_i^{(1)},b_j^{(2)},a_k^{(1)},a_l^{(2)})\} .  
\label{veridm2}
\ee
Again, null-vectors appear in the r.h.s. of the (\ref{veridm2}) due to the constraints:
\be
Z_1^2=Z_2   \virg  Z_1Z_2=\buno
\label{con2}
\ee
and the equations of motion
\be
\delta_{6k\pm 1}Z_i=[\theta_{6k\pm 1},Z_i] \virg i=1,2  .
\label{eqsmot2}
\ee 
One can further realize that just as in the $A_1^{(1)}$ case, there is a subalgebra
consisting of the upper triangular entries of $Z_i, i=1,2$, closed under the action of 
the gauge transformations  $\delta_{6k\pm 1}$. The constraints (\ref{con2}) 
and (\ref{eqsmot2}) are consistent with such reduction giving a closed subalgebra of null vectors.
The first non-trivial examples are :
\ba
level 3 &:&  b_3^{(2)}-\dx b_2^{(1)}=0 ; \nonumber \\
level 4 &:&  b_4^{(2)}- (b_2^{(1)})^2 + \frac{2}{3} \dx b_3^{(2)} =0 ; \nonumber \\
level 6 &:&  b_6^{(1)} - 2 b_6^{(2)} + 2 b_2^{(1)}b_4^{(2)} + (b_3^{(2)})^2 =0 ; \nonumber \\
          &2&b_6^{(1)} -  b_6^{(2)} + 2 \dx b_5^{(1)} + \frac{1}{2} \dx^2 b_4^{(2)} + b_2^{(1)}b_4^{(2)} - (b_2^{(1)})^3 + (b_3^{(2)})^2 =0 . 
\ea
Therefore, in order to obtain the true spectrum of the family of the identity
(i.e. the $A_2^{(2)}$-KdV spectrum), one has to factor out, from the 
freely generated Verma module
\be
{\cV}^{KdV}_{{\bf 0}}=\{\delta_{6k_1+1}\dots \delta_{6k_M+1}\delta_{6l_1-1}\dots \delta_{6l_N-1}
{\cP}(b_i^{(1)},b_j^{(2)})\} ,  
\label{veridk2}
\ee
the whole set of the null-vectors ${\cN}^{KdV}_{{\bf 0}}$, i.e. 
\be
[{\bf 0}]={\cV}^{KdV}_0 / {\cN}^{KdV}_{{\bf 0}} .
\ee
Let us turn to the classical limit of the primary fields $e^{m\phi}, m=0,1,2,3,\dots$  . 
By virtue of the above reasoning  we conjecture for their Verma modules the expression
\be
{\cV}^{mKdV}_{{\bf 0}}=\{\delta_{6k_1+1}\dots \delta_{6k_M+1}\delta_{6l_1-1}\dots \delta_{6l_N-1}
[{\cP}(b_i^{(1)},b_j^{(2)})e^{m\phi}]\} .  
\label{vermk2}
\ee
Again, we have to add to the null-vectors coming from (\ref{con2}) and (\ref{eqsmot2}) 
the new ones coming
from (repeated) application  of the equations of motion of the power $T^m(x;\la)$ 
\be
\delta_{6k\pm 1}T^m=\sum_{j=1}^mT^j\theta_{6k\pm 1}T^{m-j}
\label{eqmT^m2}
\ee 
obtaining the whole set of null-vectors ${\cN}^{KdV}_{{\bf m}}$. 
The first non-trivial examples of these additional null-vectors are given by:
\ba
level &2&: \spz (\dx^2 + 3b_2^{(1)}) e^{\phi} =0  \nonumber \\
level &3&: \spz (\dx^3 - 6 b_3^{(2)} + 12 \dx b_2^{(1)}) e^{2\phi} =0  \nonumber \\
level &4&: \spz (\dx^4 + \frac{135}{2} (b_2^{(1)})^2 + 30 \dx^2 b_2^{(1)} - \frac{27}{2} b_4^{(2)} - 30 \dx b_3^{(2)} ) e^{3\phi} =0 
\ea
where the operator $\dx$ acts on all the fields to its right. 
As a result, the ({\it conformal}) family of the primary field  $e^{m\phi}, m=0,1,2,3,\dots$ 
is conjectured to be in this case
\be
[{\bf m}]={\cV}^{KdV}_m / {\cN}^{KdV}_{{\bf m}}   .
\label{conmk2}
\ee

\section{Conclusions }

In conclusion, we presented here an alternative approach to the description of the spectrum of the local fields in the classical limit of the 2D Integrable Field Theories. It is essentially a variant of the Classical Inverse
 Scattering Method and is based on the so called Dressing Symmetries. It turns out that a number of constraints or classical null vectors appear naturaly in the context of this approach. We presented a systematic method for their derivation. These are due to (succesive application of) the constraints and equations of motion obeyed by the asymptotic expansion of our basic objects - the transfer matrix $T(x;\lambda)$ 
and the resolvent $Z(x;\lambda)$.

Instead, the regular case is connected with a symmetry of Poisson--Lie type. The construction of the spectrum employing this symmetry is an interesting problem, probably connected
to the BRST prescription of \cite{F}.
 Another related problem is the asymptotic counterpart of the above symmetry arising from the dressing of the Chevalley generators not belonging to the Cartan subalgebra. It turns out that the 
resulting generators do not commute with the KdV ones but rather close a kind of spectrum generating algebra. These problems will be treated in detail in a forthcoming paper \cite{35}.

An important question is the quantization of the classical constructions presented above. We claim, 
in compliance with \cite{BLZ1}, that a consistent quantization procedure consists in substituting the underlying Kac-Moody algebra with the corresponding quantum one and the fundamental mKdV field $\phi (x)$ with 
the Feigin-Fuchs free scalar field. As a consequence our regular Poisson--Lie symmetry 
turns into the well known quantum symmetry of CFT. The asymptotic construction instead should lead to a 
generalization of the results of \cite{BBS} for cases beyond $A^{(1)}_1$.

Another open problem is the extending of our constructions out of criticality. Again, in accordance with 
\cite{BLZ1}, we suggest for the off-critical transfer matrix 
${\bf T}(x,\bar x;\lambda ;\mu)=\bar T(\bar x;\mu/\lambda) T(x;\lambda)$ where:
\be
\bar T(\bar x,\la) ={\cal P}
\exp\lt \la \int_0^{\bar x}dy
(e^{-2{\bar \phi}(y)} E+ e^{2{\bar \phi}(y)} F ) \rt e^{H{\bar \phi}(\bar x)}
\label{Tba}
\ee
and correspondingly for ${\bf Z}(x,\bar x; \mu ;\lambda)$. It is clear that in this case some of their properties, in particular 
the equations of motion, will change. One can try to overgo this problem by
following again an approach close to \cite{SS}. In particular we claim that some modification 
of the construction based on the extrinsic geometry of surfaces presented there will prove useful. 
We shall return to these problems elsewhere \cite{36}.

{\bf Acknowledgments} - We thank P.E.Dorey, M.
de Innocentis, A. Ganchev, V. Petkova, A. Sagnotti, G. Sotkov for discussions and interest in this work. D.F. thanks 
the EC Commission for financial support through TMR Contract ERBFMRXCT960012 and acknowledges I.N.F.N.-Rome (II Univ.) 
and the Laboratoire de Physique Mathematique (Univ. de Montpellier II) 
for the kind hospitality during part of this work. The same does M.S. with I.N.F.N.-Bologna.

\end{document}